\documentclass[runningheads]{llncs}
\usepackage{graphicx}
\usepackage{amsmath}
\usepackage{color}

\begin{document}
\title{Fitting a Directional Microstructure Model to Diffusion-Relaxation MRI Data with Self-Supervised Machine Learning}

\titlerunning{Self-supervised directional microstructure modelling}

\author{Jason P. Lim\inst{1}\and
Stefano B. Blumberg\inst{1,2} \and
Neil Narayan\inst{1} \and
Sean C. Epstein\inst{1} \and
Daniel C. Alexander\inst{1} \and
Marco Palombo \inst{1,3,4} \and
Paddy J. Slator \inst{1}}

\authorrunning{J. Lim et al.}

\institute{Centre for Medical Image Computing, Department of Computer Science, University College London, London, UK \and
Centre for Artificial Intelligence, Department of Computer Science, University College London, UK \and
Cardiff University Brain Research Imaging Centre (CUBRIC), School of Psychology, Cardiff University, Cardiff, UK \and
School of Computer Science and Informatics, Cardiff University, Cardiff, UK \\
\email{p.slator@ucl.ac.uk}}

\maketitle              

\begin{abstract}
Machine learning is a powerful approach for fitting microstructural models to diffusion MRI data. 
Early machine learning microstructure imaging implementations trained regressors to estimate model parameters in a supervised way, using synthetic training data with known ground truth. However, a drawback of this approach is that the choice of training data impacts fitted parameter values.
Self-supervised learning is emerging as an attractive alternative to supervised learning in this context.
Thus far, both supervised and self-supervised learning have typically been applied to isotropic models, such as intravoxel incoherent motion (IVIM), as opposed to models where the directionality of anisotropic structures is also estimated. 
In this paper, we demonstrate self-supervised machine learning model fitting for a directional microstructural model.
In particular, we fit a combined T1-ball-stick model to the multidimensional diffusion (MUDI) challenge diffusion-relaxation dataset.
Our self-supervised approach shows clear improvements in parameter estimation and computational time, for both simulated and in-vivo brain data, compared to standard non-linear least squares fitting.
Code for the artificial neural net constructed for this study is available for public use from the following GitHub repository: https://github.com/jplte/deep-T1-ball-stick.

\end{abstract}

\section{Introduction}
Microstructure imaging aims to quantify features of the tissue microstructure from in-vivo MRI \cite{alexanderImagingBrainMicrostructure2019}. Historically, microstructure imaging utilised diffusion MRI (dMRI) data. Recently, combined diffusion-relaxation MRI - where relaxation-encoding parameters such as inversion time (TI) and echo time (TE) are varied alongside diffusion-encoding parameters such as b-value and gradient direction - has been emerging as an extension \cite{slatorCombinedDiffusionrelaxometryMicrostructure2021}.
The typical approach to estimating tissue microstructure from such diffusion or diffusion-relaxation data is multi-compartment modelling, which utilises signal models comprising linear combinations of multiple compartments - such as balls, sticks, zeppelins, and spheres - each representing a distinct tissue geometry\cite{panagiotakiCompartmentModelsDiffusion2012}.  

Multi-compartment microstructure models are usually fit to the data with non-linear least squares (NLLS) algorithms. However, these can be computationally expensive and are prone to local minima, necessitating grid searches or parameter cascading \cite{fickDmipyToolboxDiffusion2019} to seek global minima. Machine learning  is a powerful alternative. Thus far, most machine learning microstructure model fitting approaches have used supervised learning \cite{golkovQSpaceDeepLearning2016,nedjati-gilaniMachineLearningBased2017,liFastRobustDiffusion2019,palomboSANDICompartmentbasedModel2020,palomboJointEstimationRelaxation2021,diaoParameterEstimationWMTIWatson2022,kerkelaImprovedReproducibilityDiffusion2022}. However, a crucial limitation is that the distribution of training data significantly affects fitted parameters \cite{gyoriTrainingDataDistribution2022,epsteinChoiceTrainingLabel2022}. It has also proved difficult to estimate directional parameters, such as fibre direction, with existing machine learning methods instead directly estimating rotationally invariant parameters, such as mean diffusivity, fractional anisotropy, mean kurtosis, and orientation dispersion.
This may be due to the difficulty of constructing a  training dataset that adequately samples the high-dimensional parameter space, and/or complications due to the periodicity of angular parameters.

Self-supervised  (sometimes imprecisely called unsupervised in the microstructure imaging context)  learning is an alternative with the potential to address these limitations.
Self-supervised algorithms learn feature representations from the input data by inferring supervisory constraints from the data itself.
For microstructure imaging, self-supervised learning has been implemented with voxelwise fully connected artificial neural networks (ANNs).
However, thus far self-supervised microstructure imaging has been limited to isotropic models\cite{grussuDeepLearningModel2021,dealmeidamartinsNeuralNetworksParameter2021}, including many intravoxel incoherent motion (IVIM) MRI examples\cite{barbieriDeepLearningHow2020,kaandorpImprovedUnsupervisedPhysicsinformed2021,vasylechkoSelfsupervisedIVIMDWI2021,zhouUnsupervisedDeepLearning2022,epsteinChoiceTrainingLabel2022}.
To our knowledge, self-supervised model fitting has not yet been demonstrated for directional microstructural models.

In this paper, we fit an extended T1-ball-stick model to diffusion-relaxation MRI data using self-supervised machine learning and demonstrate several advantages of this approach over classical NLLS, such as higher precision and faster computational time.

\section{Methods}
\subsection{Microstructure model} 
As this is a first attempt at fitting directional multi-compartment models with self-supervised learning, we choose a simple model – the ball-stick model first proposed by Behrens et al. \cite{behrensCharacterizationPropagationUncertainty2003}. According to the ball-stick model, the expression for the normalized signal decay is
\begin{equation}
    S(b, \mathbf{g}) = f \exp\left(-b \lambda_{||} (\mathbf{g}.\mathbf{n})\right)
    + (1-f) \exp\left(-b\lambda_{iso}\right) 
\end{equation}
where $b$ is the b-value, $\mathbf{g}$ is the gradient direction,  $\lambda_{||}$ and $\lambda_{iso}$ are the parallel and isotropic diffusivities of the stick and ball respectively, and $\mathbf{n}$ is the stick orientation, which we parameterise using polar coordinates.
The relationship between Cartesian and polar coordinates is $n=[\sin{\theta} \cos{\phi},\sin{\theta} \sin{\phi},\cos{\theta}]$
where  $\phi \in [0,\pi]$ and $\theta \in [-\pi,\pi]$.

We extend the ball-stick model to account for T1 relaxation time, by assuming the ball and stick compartments have separate T1 times, represented by $T1_{ball}$ and $T1_{stick}$ respectively. 
Note that we assume a single T2 for both compartments, so the volume fraction $f$ will be affected by the T2 of each compartment. 
Given a combined T1 inversion recovery \cite{bydderMRImagingClinical1985} and diffusion MRI experiment, where inversion time (TI), b-value and gradient direction are simultaneously varied, we can fit the following T1-ball-stick equation 
\begin{align}
    S&(b,g,T_I,T_R) = f \exp\left(-b \lambda_{||} (\mathbf{g}.\mathbf{n})\right) \left| 1 - 2\exp\left(-\frac{T_I}{T1_{stick}}\right) \exp\left(-\frac{T_R}{T1_{stick}}\right) \right| \notag \\
   &+(1-f) \exp\left(-b\lambda_{iso}\right) \left| 1 - 2\exp\left(-\frac{T_I}{T1_{ball}}\right) \exp\left(-\frac{T_R}{T1_{ball}}\right) \right|
   \label{eq:T1-ball-stick}
\end{align}
In this work, we first fit this model to combined T1-diffusion data with standard NLLS, then demonstrate self-supervised fitting with an ANN. We first describe the data, then the model fitting techniques. 

\subsection{Combined T1-diffusion in vivo data}
We utilise in-vivo data from 5 healthy volunteers (3 F, 2 M, age=19–46 years), acquired from the 2019 multidimensional diffusion (MUDI) challenge \cite{pizzolatoAcquiringPredictingMultidimensional2020}. The acquisition sequence comprises simultaneous diffusion, inversion recovery (giving T1 contrast), and multi-echo gradient echo  (giving T2* contrast) measurements. We chose to ignore the subsection of the data that is sensitive to T2*  by only included signals captured with the lowest  echo time (80 ms). This is since the two higher TEs have very low signal intensity and the 3 TEs have a small range, and thus there is limited T2* information in the data. Our subsequent description hence only refers to the subsection of the data with TE = 80 ms.

The datasets were obtained using a clinical 3T Philips Achieva scanner (Best, Netherlands) with a 32-channel adult headcoil. Each scan includes 416 volumes distributed over five b-shells, b $\in$ $\{0, 500, 1000, 2000,3000\}$ s/mm$^2$, with 16 uniformly spread directions, and 28 inversion times (TI) $\in$ [20, 7322] ms. For all datasets, the following parameters were fixed: repetition time TR=7.5 s, resolution=2.5 mm isotropic, FOV=220×230×140 mm, SENSE=1.9, halfscan=0.7, multiband factor 2, total acquisition time 52 min (including preparation time). 

\begin{figure}\centering
\includegraphics[width=0.9\textwidth]{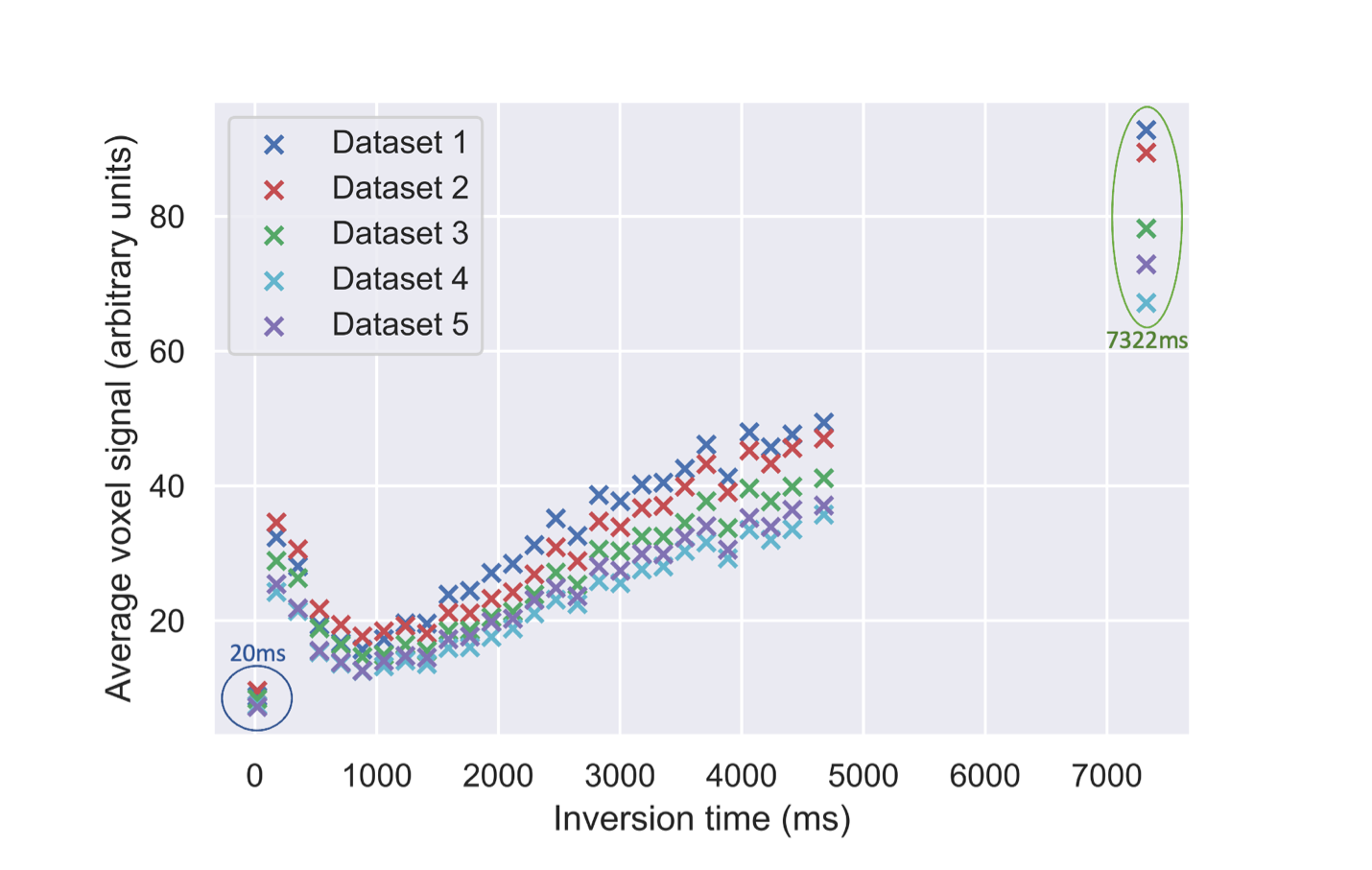}
\caption{Average signal of all b=0 voxels within the brain mask at each inversion time. Note the outlying signals at the smallest (20ms, blue) and largest (7322ms, green) inversion times. We removed these from the data before fitting. } \label{fig:data-outliers}
\end{figure}

The MUDI data has already undergone standard pre-processing,  see \cite{pizzolatoAcquiringPredictingMultidimensional2020} for full details. Upon inspection, we noted that the lowest (20 ms) and highest (7322 ms) inversion times, which comprise 7.14\% of the data, were clearly dominated by noise and/or artifacts (see Figure \ref{fig:data-outliers}). We therefore removed them from the data prior to model fitting, leaving 416 MRI volumes in total. After removing these TIs, the dataset contains 26 TIs $\in$ [176, 4673] ms. 

We normalised each voxel’s data independently, by dividing by the signal generated from the b=0 volume with the highest TI, i.e. the volume with the highest expected signal. We then removed all background voxels using the brain mask provided with the MUDI data.

\subsection{Simulated data}
In-vivo MRI data does not have ground truth tissue-related parameters values, making it hard to quantitatively assess the accuracy of model fitting. We thus simulated 100,000 synthetic signals using the T1-ball-stick model signal equation (Equation \eqref{eq:T1-ball-stick}). We used the same 416 acquisition parameters (b-values, gradient directions and inversion times) as our reduced MUDI dataset. The ground truth values of $\lambda_{||}$, $\lambda_{iso}$, $T1_{ball}$ and $T1_{stick}$ were sampled randomly from physically-plausible ranges (Table \ref{table:parameter-ranges}). Note that we choose  units so that parameter values are close to 1; this prevents having to normalise parameter values before training neural networks. Complex Gaussian noise was added to simulate the Rician distribution of noisy MRI data \cite{koayAnalyticallyExactCorrection2006}.
\begin{table}\centering
\caption{Constraints on T1-ball-stick parameters for simulating data and model fitting.}\label{table:parameter-ranges}
\begin{tabular}{|c|c|c|}
\hline
Parameter &  Minimum & Maximum\\
\hline
$\lambda_{iso}$ &  0.1 \textmu m$^2$/ms & 3.0 \textmu m$^2$/ms \\
$\lambda_{||}$ &  0.1 \textmu m$^2$/ms & 3.0 \textmu m$^2$/ms \\
$f$ & 0 & 1 \\
$T1_{ball}$ &  0.01 s  & 5 s \\
$T1_{stick}$ &  0.01 s  & 5 s \\
$\theta$ & 0  & $\pi$ \\
$\phi$ & $-\pi$ & $\pi$ \\
\hline
\end{tabular}
\end{table}

\subsection{Non-linear least squares fitting}
The modified T1-ball-stick model (Equation \eqref{eq:T1-ball-stick}) was fit with non-linear least squares by modifying the open source diffusion microstructure imaging in python (dmipy) toolbox \cite{fickDmipyToolboxDiffusion2019}, with parameter constraints as in Table \ref{table:parameter-ranges}. Specifically, we used the ``brute2fine" function, which uses a brute force grid search followed by non-linear optimisation.

\subsection{Self-supervised model fitting}
\begin{figure}
\includegraphics[width=\textwidth]{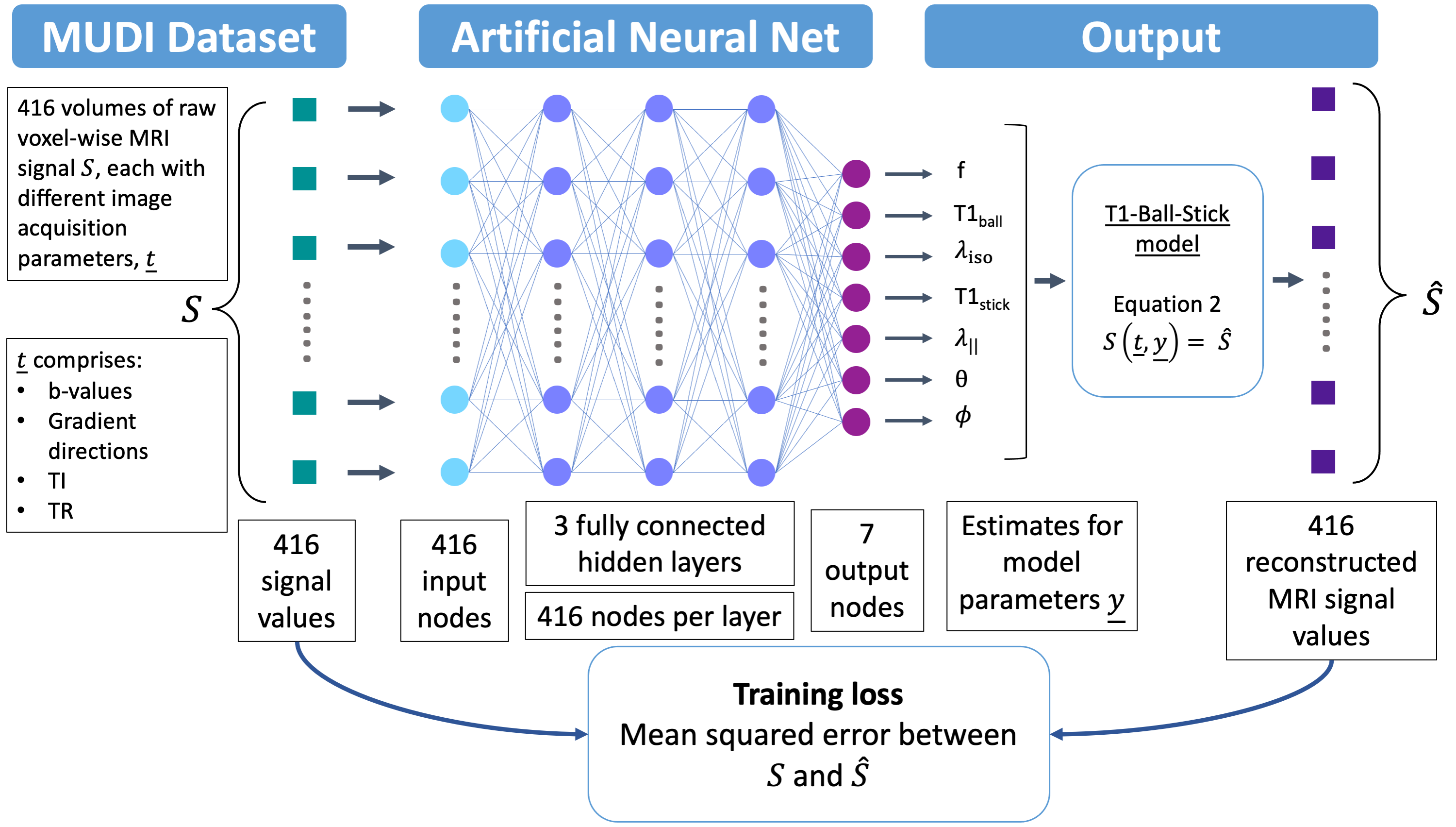}
\caption{Our ANN for T1-ball-stick fitting and the flow of data through it.} \label{fig:ANNdiagram}
\end{figure}

We built an ANN to generate estimates for the T1-ball-stick parameters. The network comprises an input layer, 3 fully connected hidden layers and an output layer, (see Figure \ref{fig:ANNdiagram}). The input layer and hidden layers each have 416 nodes - mirroring the 416 MRI volumes.

The final layer has 7 output neurons, one for each parameter of interest. 
The normalised signal from a single voxel of the MRI data, $S$, which comprises 416 measurements, is fed into the input layer and passed through the ANN. The output layer is fed forward into the T1-ball-stick model equation, giving a synthetic signal $\hat{S}$. Training loss is the mean squared error between input ($S$) and synthetic ($\hat{S}$) MRI signals across all voxels passed through the ANN.

We implemented the ANN on Python 3.9.5 using PyTorch21 1.10.0. 
For both simulated and in-vivo data, we used the Adam optimiser\cite{kingmaAdamMethodStochastic2017} with learning rate 0.0001, batch size 128, and dropout\cite{srivastavaDropoutSimpleWay2014} with rate 0.5. 
Parameter constraints (Table \ref{table:parameter-ranges}) were imposed using PyTorch’s clamp feature, which converts any value outside the bounds to the value closest to it within the boundary. 
Following\cite{barbieriDeepLearningHow2020}, we trained the network with patience 10, i.e. until there 10 consecutive epochs without loss improvement.
As the self-supervised approach estimates T1-ball-stick parameters directly from the data, we didn't use a train-test split. Instead, the ANN was trained on each dataset separately.

\section{Results}
\subsection{Simulated data}
\begin{figure}
    \centering
    \includegraphics[width=0.94\textwidth]{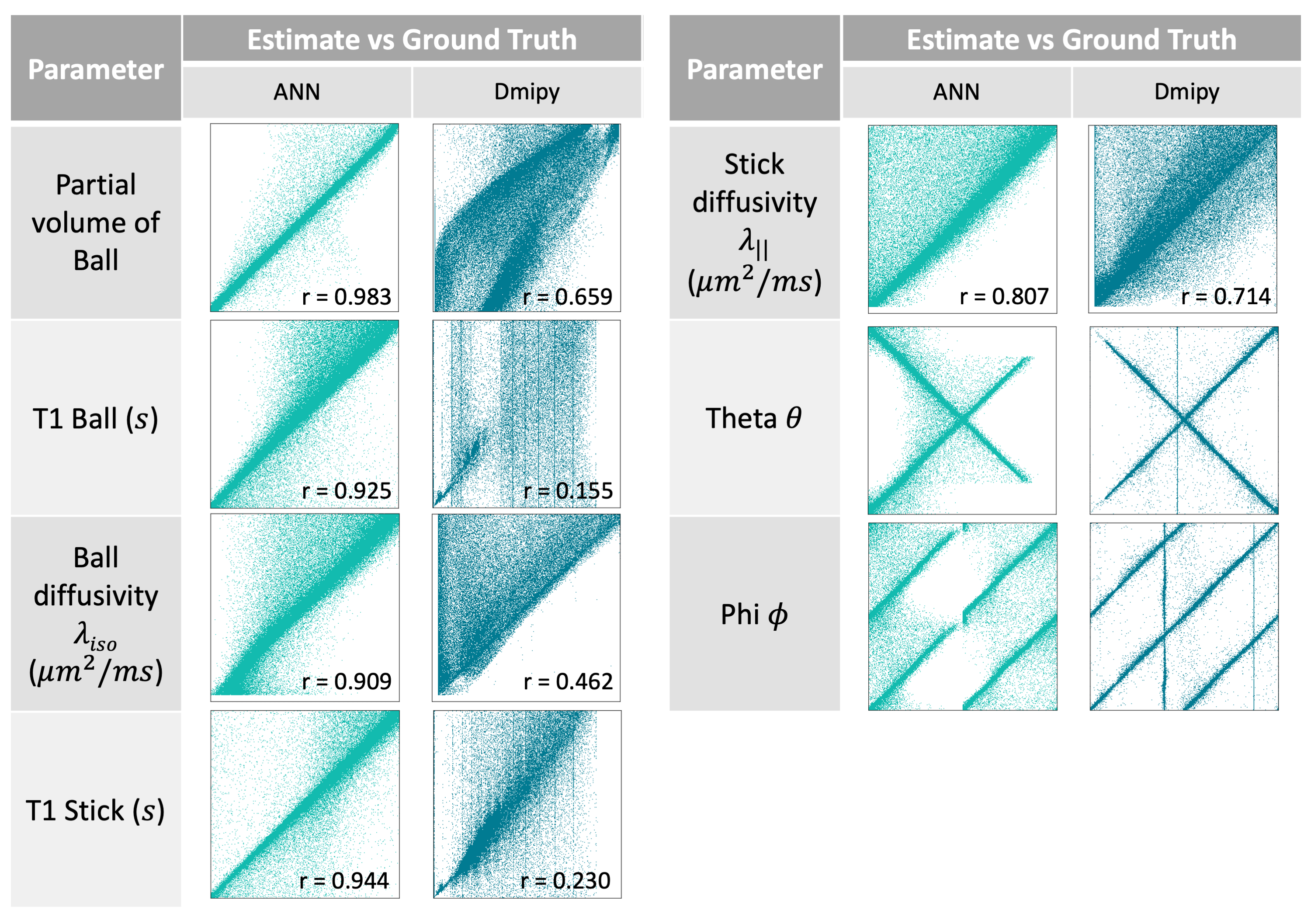}
    \caption{NLLS and ANN T1-ball-stick estimates against ground truth values, after fitting the  model to simulated signals. r denotes the Pearson correlation coefficient. }
    \label{fig:sims}
\end{figure}

The T1-ball-stick model was successfully fit to the simulated signals using both the ANN and NLLS. Scatter plots and Pearson correlation coefficients of parameter estimates against  ground truth values are shown in Figure \ref{fig:sims} (note that we don't report correlations for $\theta$ and $\phi$ as the values are confounded by the periodicity of these angular parameters).
Correlation coefficients for the ANN fits are higher for all model parameters, with coefficients above 0.9 for all parameters except $\lambda_{||}$. NLLS correlation coefficients for T1 relaxation time are particularly low. 

\subsection{Real data}
\begin{figure}
    \centering
    \includegraphics[width=1.2\textwidth]{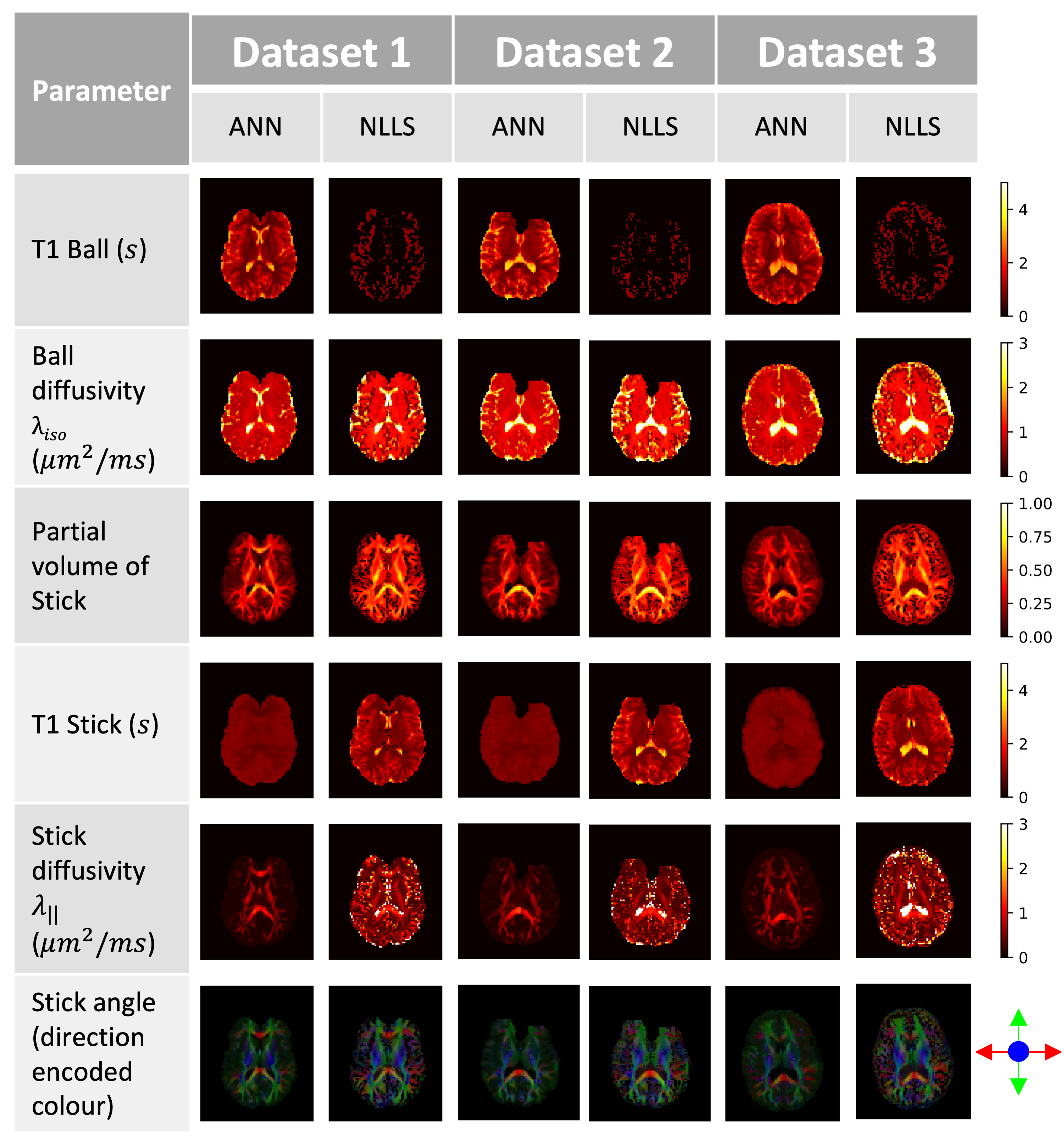}
    \caption{Parameter maps for all models fit to the first 3 MUDI datasets using both Dmipy and ANN methods. Each parameter map is a cross-sectional view of the brain. The maps presented are generated from the middle Z-axis slice.}
    \label{fig:real-maps1}
\end{figure}

\begin{figure}
    \centering
    \includegraphics[width=\textwidth]{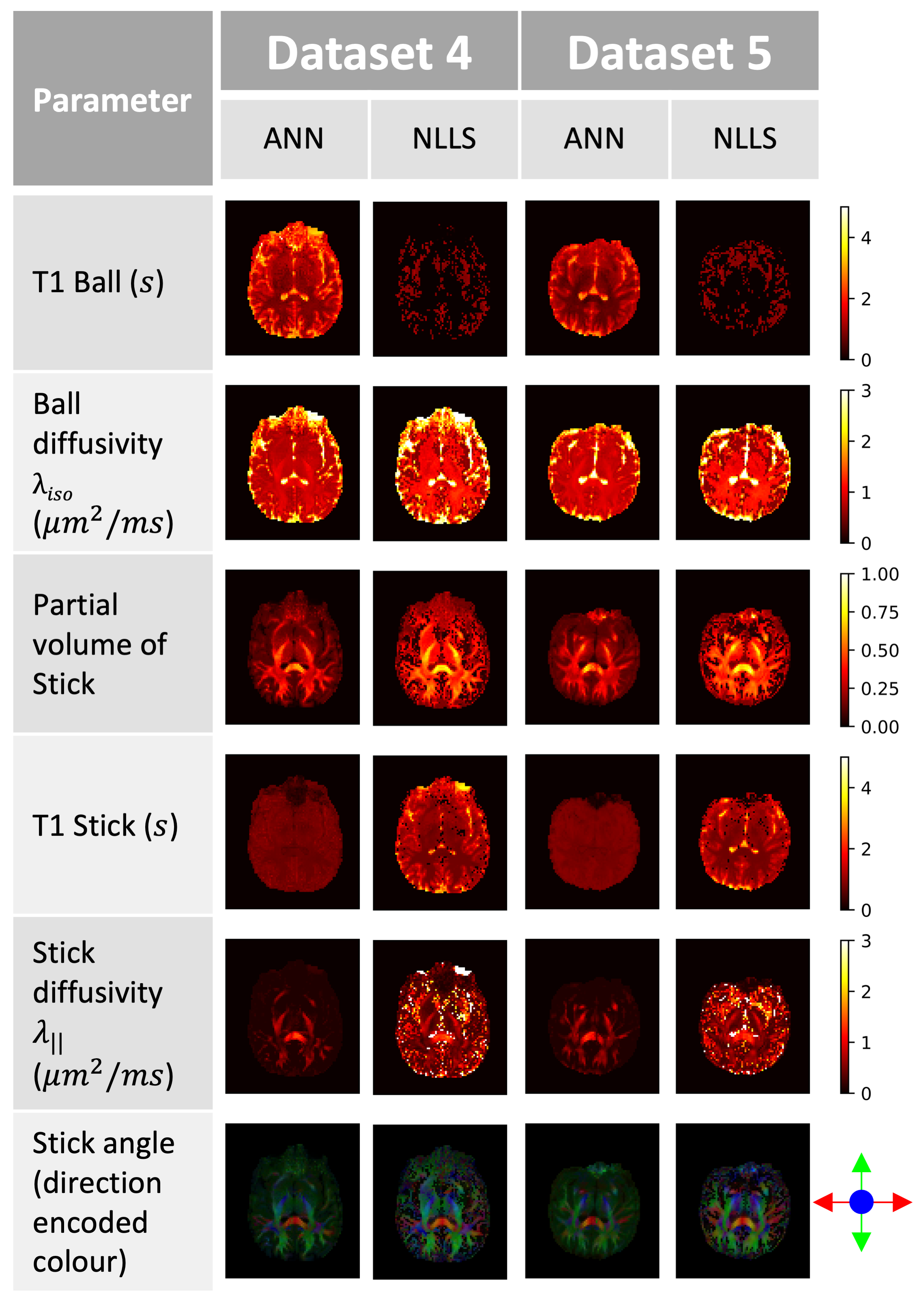}
    \caption{As Figure \ref{fig:real-maps1} but for the remaining 2 MUDI datasets. }
    \label{fig:real-maps2}
\end{figure}

Figures \ref{fig:real-maps1} and \ref{fig:real-maps2} show T1-ball-stick model fits for all MUDI subjects. ANN parameter maps are qualitatively less noisy and show more anatomically plausible contrast than NLLS. This is particularly clear for $T1_{ball}$, $\lambda_{||}$ and direction encoded colour (DEC) maps.
All 5 MUDI subjects showed similar trends. ANN inferred  higher and lower  values  than NLLS for $T1_{ball}$ and $T1_{stick}$ respectively. The ANN clearly shows highest  $\lambda_{||}$ values in the corpus callosum, while the Dmipy fit has high $\lambda_{||}$ values in many places. $\lambda_{iso}$ maps are generally similar across both methods. 


Average time taken for ANN fits on real data was 1966 s, compared to 8833 s in Dmipy, meaning ANN was 77.25\% faster than NLLS on average. All model fits were performed on a 2017 Macbook Pro’s central processing unit (3.1 GHz Dual-Core Intel I5-7267U).

\section{Discussion}
This study demonstrates self-supervised microstructure imaging for a combined T1-ball-stick model.  Our ANN approach is faster and more precise that conventional NLLS model fitting. In the ANN model fits, parametric maps show plausible estimates for both diffusivity ($\lambda_{||}$, $\lambda_{iso}$) and relaxation ($T1_{ball}$ and $T1_{stick}$), whereas some NLLS maps, particularly $T1_{ball}$, show dubious contrast.

The ANN model fits potentially reveal more accurate tissue information than NLLS counterparts. 
Throughout the brain, white matter T1 times are expected to be around 0.7-0.9 s \cite{bojorquezWhatAreNormal2017}. The ANN estimates fall within this range – the $T1_{stick}$ voxels displayed in Figures \ref{fig:real-maps1} and \ref{fig:real-maps2} average to 0.87 s - while NLLS estimates are higher, with some regions reaching 4 s. T1 times in the CSF are expected to be around 4 s \cite{bojorquezWhatAreNormal2017}, which is reflected in the ANN $T1_{ball}$ estimate, but not in the NLLS estimate, where it is approximately 0 s (Figures \ref{fig:real-maps1} and \ref{fig:real-maps2}).

These observations match those in the simulations (Figure \ref{fig:sims}), with NLLS correlation being extremely poor in both T1 times. In line with our qualitative analysis of MUDI data fits, Figure  \ref{fig:sims} shows that the ANN outperforms Dmipy in every parameter, except  potentially for $\theta$ and $\phi$, whose correlations are not straightforward to quantify due to their periodicity. 
Direction encoded colour (DEC) maps are less noisy for ANN fits, with less visible noise, and are hence potentially more useful for tractography.

The ANN approach was faster in all datasets, with a 77\% improvement in time on average. This will hopefully improve the feasibility of utilising similar modelling in clinical scenarios. However, the ANN would still require retraining for every new dataset. A possible next step would be to explore the viability of using multiple datasets to train an ANN to be generalisable to unseen data, then fine-tuning the network on each new dataset, effectively combining advantages of supervised and self-supervised learning, as recently demonstrated by Epstein et al. \cite{epsteinChoiceTrainingLabel2022}. If successful, this could significantly reduce time taken to generate parameter estimates for new patients, but possibly at the cost of accuracy.

Whilst the ANN outperforms NLLS in our experiments, we applied NLLS ``out of the box" without focusing on improving the fitting. {\color{black}The NLLS fits could be improved - e.g. the vertical lines in Figure \ref{fig:sims} are likely local minima.  }. In future, we could use a larger grid in the grid search stage, although this can quickly lead to infeasible computational times, or initialise the NLLS fit from ``reasonable" parameter values. Whilst these would likely improve the NLLS fits, the fact that our self-supervised ANN bypasses these ad-hoc tuning steps presents a significant advantage. {\color{black}We also only compare parameter estimates with the ground truth using correlation coefficients, effectively merging bias and variance. In future, we could calculate bias and variance separately, and also explore tuning the training cost function towards accuracy or precision depending on the application. Ball-stick is a very simple single bundle model, in future we will explore multi-fibre models that more accurately reflect brain microstructure.}

The small sample size is a limitation of this study. In future, we can adapt our ANN to fit  standard microstructural models that only require diffusion MRI data. This would enable us to test self-supervised learning against NLLS, and quantify test-retest repeatability, on large open source datasets. Additionally, the 5 datasets are all from healthy patients with normal physiology. Hence, we are unable to judge the suitability of the T1-ball-stick model parameters as imaging biomarkers. The overall goal is to identify imaging biomarkers that can be used for diagnosis, prognosis, and monitoring of brain conditions such as stroke and dementia, so future studies should involve MRI data from patients with these conditions. This would also help us determine thresholds for model parameters to differentiate between healthy and diseased tissue.

\section{Conclusion}
We demonstrate, for the first time, self-supervised learning fitting of a directional microstructural model, T1-ball-stick. We show vastly improved performance, in terms of speed and accuracy, compared to the current standard fitting technique, NLLS. Self-supervised machine learning model fitting had only been demonstrated in a limited number of simple MC models thus far, such as IVIM \cite{barbieriDeepLearningHow2020}. This work can pave the way for self-supervised fitting of a wide range of multi-compartment microstructure models to MRI data.

\section{Acknowledgments}
This work was supported by EPSRC grants  EP/M020533/1, EP/V034537/1, and EP/R014019/1; UKRI Future Leaders Fellowship MR/T020296/2 (MP); an EPRSC and Microsoft scholarship (SBB); and
the National Institute for Health Research (NIHR) Biomedical Research Centre at University College London Hospitals NHS Foundation Trust and University College London.
The views expressed are those of the authors and not necessarily those of the NHS, the NIHR or the Department of Health.

\bibliographystyle{splncs04}
\bibliography{bibliography.bib}

\end{document}